\documentclass[aps,prl,twocolumn,showpacs,groupedaddress]{revtex4}

\usepackage{graphicx}
\usepackage{mathtools,amssymb}
\usepackage{siunitx}

\begin{document}

\preprint{}

\title{Origin of the two-dimensional electron gas at LaAlO$_{3}$/SrTiO$_{3}$ interfaces - The role of oxygen vacancies and electronic reconstruction}

\author{Z. Q. Liu$^{1,2}$}

\altaffiliation{These authors contributed equally to this work}

\author{C. J. Li$^{1,3}$}

\altaffiliation{These authors contributed equally to this work}

\author{W. M. L\"{u}$^{1}$}

\altaffiliation[Corresponding author: ]{weiming@nus.edu.sg}

\author{X. H. Huang$^{4}$}

\author{Z. Huang$^{1}$}

\author{S. W. Zeng$^{1,2}$}

\author{X. P. Qiu$^{4}$}

\author{L. S. Huang$^{5}$}

\author{A. Annadi$^{1,2}$}

\author{J. S. Chen$^{5}$}

\author{J. M. D. Coey$^{1,6}$}

\author{T. Venkatesan$^{1,2,3,4}$}

\author{Ariando$^{1,2}$}

\altaffiliation[Corresponding author: ]{ariando@nus.edu.sg}

\affiliation{$^1$NUSNNI-Nanocore, National University of Singapore, 117411 Singapore}

\affiliation{$^2$Department of Physics, National University of Singapore, 117542 Singapore}

\affiliation{$^3$National University of Singapore Graduate School for Integrative Sciences and Engineering (NGS), National University of Singapore, 117456 Singapore}

\affiliation{$^4$Department of Electrical and Computer Engineering, National University of Singapore, 117576 Singapore}

\affiliation{$^5$Department of Material Science and Engineering,
National University of Singapore, 117576 Singapore}

\affiliation{$^6$Department of Pure and Applied Physics, Trinity College, Dublin 2, Ireland}

\date{\today}

\begin{abstract}
The relative importance of atomic defects and electron transfer in explaining conductivity at the crystalline LaAlO$_{3}$/SrTiO$_{3}$ interface has been a topic of debate. Metallic interfaces with similar electronic properties produced by amorphous oxide overlayers on SrTiO$_{3}$ [Y. Chen \emph{et al}., Nano Lett. {\bf 11}, 3774 (2011); S. W. Lee \emph{et al}., Nano Lett. {\bf 12}, 4775 (2012)] have called in question the original polarization catastrophe model [N. Nakagawa \emph{et al}., Nature Mater. {\bf 5}, 204 (2006)]. We resolve the issue by a comprehensive comparison of (100)-oriented SrTiO$_{3}$ substrates with crystalline and amorphous overlayers of LaAlO$_{3}$ of different thicknesses prepared under different oxygen pressures. For both types of overlayers, there is a critical thickness for the appearance of conductivity, but its value is always 4 unit cells ($\sim$1.6 nm) for the oxygen-annealed crystalline case, whereas in the amorphous case the critical thickness could be varied in the range 0.5 to 6 nm according to the deposition conditions. Subsequent ion milling of the overlayer restored the insulating state for the oxygen-annealed crystalline heterostructures but not for the amorphous ones. Oxygen post-annealing removes the oxygen vacancies, and the interfaces become insulating in the amorphous case, but the interfaces with a crystalline overlayer remain conducting with reduced carrier density. These results demonstrate that oxygen vacancies are the dominant source of mobile carriers when the LaAlO$_{3}$ overlayer is amorphous, while both oxygen vacancies and polarization catastrophe contribute to the interface conductivity in unannealed crystalline LaAlO$_{3}$/SrTiO$_{3}$ heterostructures, and the polarization catastrophe alone accounts for the conductivity in oxygen-annealed crystalline LaAlO$_{3}$/SrTiO$_{3}$ heterostructures. Furthermore, it was found that the crystallinity of the LaAlO$_{3}$ layer is crucial for the polarization catastrophe mechanism in the case of crystalline LaAlO$_{3}$ overlayers.
\end{abstract}



\maketitle

The two-dimensional electron gas (2DEG) appearing at the interface between the band insulators LaAlO$_{3}$ (LAO) and SrTiO$_{3}$ (STO) has attracted much attention since its discovery by Ohtomo and Hwang [1]. It has stimulated a substantial body of experimental and theoretical work [2-25], but, its origin is still controversial [26]. Three different mechanisms have been proposed. First is interface electronic reconstruction to avoid the polarization catastrophe induced by the discontinuity at the interface between polar LAO and nonpolar STO [2-4]. Second is doping by thermal interdiffusion of Ti/Al or La/Sr atoms at the interface [13]. A third possible mechanism is creation of oxygen vacancies in STO substrates during the deposition process [9-11,27,28]. Oxygen vacancies are known to introduce a shallow intragap donor level close to the conduction band of STO [29], and their action may be specific to this one substrate. The thermal interdiffusion mechanism was discounted in recent work [25], which studied the effect of a mixed interface layer. It is also in conflict with the experimental results that \emph{p}-type LAO/STO interfaces [1] and interfaces created by growing STO films on LAO are insulating [30].

In 2007, Shibuya \emph{et al}. [28] associated the metallic conductivity at interfaces between room temperature-deposited amorphous CaHfO$_{3}$ films and STO single crystal substrates with the bombardment of STO substrates by the plume during the pulsed laser deposition process. Later, Chen \emph{et al}. [31] demonstrated metallic interfaces between STO substrates and various amorphous oxide overlayers including LAO, STO and yittria-stablized zirconia thin films fabricated by pulsed laser deposition. The origin of the 2DEG in such amorphous heterostructures was attributed to formation of oxygen vacancies at the surface of the STO. Moreover, metallic interfaces between Al-based amorphous oxides and STO substrates have also been realized by other less energetic deposition techniques such as atomic layer deposition [32] and electron beam evaporation [33]. The electronic properties of STO-based amorphous heterostructures [31,32] are, to some extent, similar to those of crystalline LAO/STO heterostructures [7,8], including the metallicity accompanied by the presence of Ti$^{3+}$ ions and a sharp metal-insulator transition as a function of overlayer thickness. These results call into question the polarization catastrophe model.

\begin{figure*}
\includegraphics[width=6.75 in]{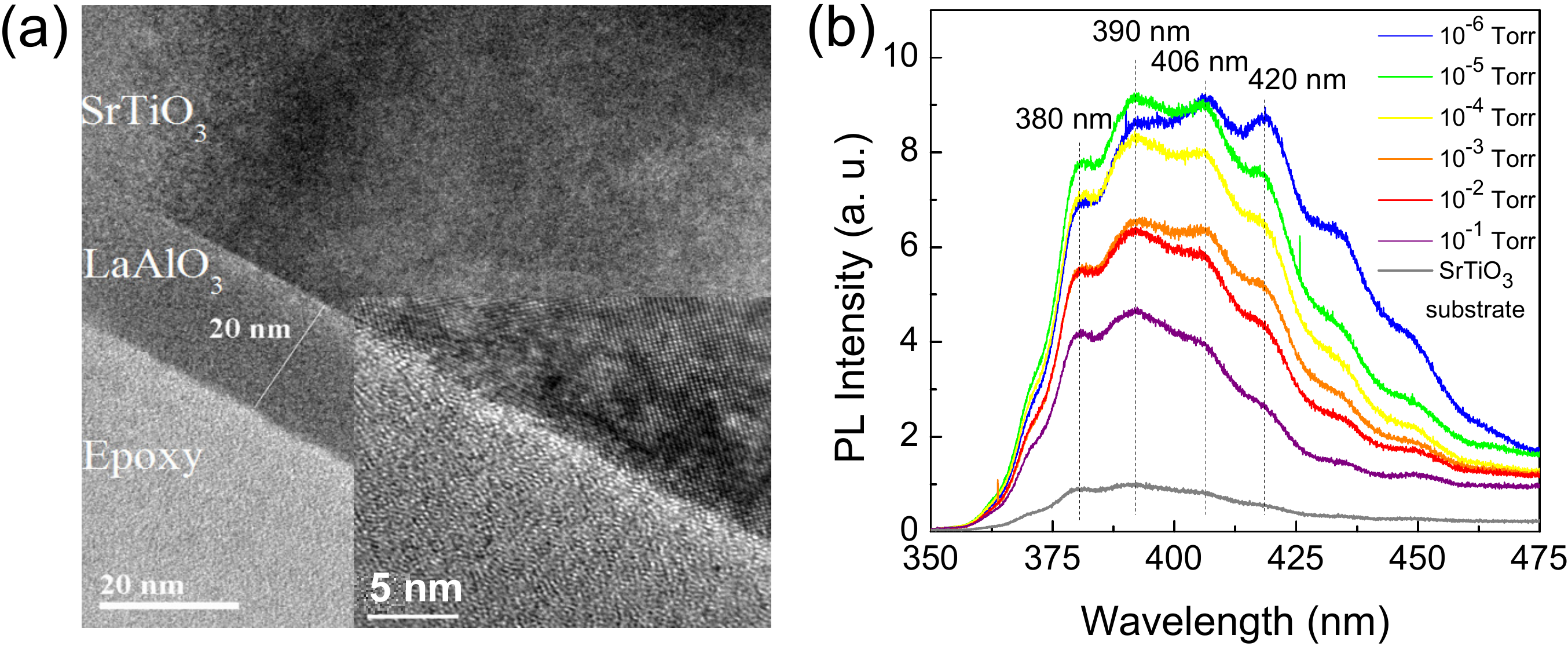}
\caption{\label{fig1} Structural and optical properties of amorphous LaAlO$_{3}$/SrTiO$_{3}$ (LAO/STO) heterostructures. (a) Transmission electron microscopy image of a LAO film deposited on an untreated STO substrate at room temperature and 10$^{-6}$ Torr oxygen pressure. Inset: zoom-in image of an interface region. (b) Room-temperature photoluminescence (PL) spectra of an as-received STO substrate and 20 nm amorphous LAO films deposited on untreated STO substrates at different oxygen partial pressure ranging from 10$^{-1}$ to 10$^{-6}$ Torr.}
\end{figure*}

In this paper, we present a detailed comparison of amorphous and crystalline LAO/STO heterostructures based on electrical and optical measurements.  By comparing the electronic properties of these two interfaces, we are able to distinguish the two different mechanisms mainly responsible for the 2DEG observed in amorphous and crystalline LAO/STO heterostructures.

Amorphous LAO films were deposited from a single crystal LAO target on untreated (100)-oriented STO substrates by pulsed laser deposition (KrF laser $\lambda$ = 248 nm) at room temperature and different oxygen partial pressures. Crystalline LAO films were fabricated on TiO$_{2}$-terminated (100)-oriented STO substrates at 750 $^{\circ}$C in 10$^{-3}$ Torr oxygen partial pressure. During deposition, the repetition rate of the laser was kept at 5 Hz and the laser fluence was fixed at 1.3 J/cm$^{2}$. The deposition rate of amorphous LAO films was calibrated by transmission electron microscopy measurements and the growth of crystalline LAO films was monitored in situ by reflection high energy electron diffraction (RHEED). Electrical contacts onto $5\times5$ mm$^{2}$ samples were made with Al wires using wire bonding and electrical measurements were performed in a Quantum Design physical property measurement system. While the sheet resistance and Hall effect of all LAO/STO heterostructures was measured in the van der Pauw geometry, the magnetoresistance (MR) measurements were performed in the four-probe linear geometry. The Ar-milling experiments were performed in an INTEL VAC Ion-Beam Milling System. Ar$^{+}$ ions beam accelerated at 200 V at 4 ml/min and irradiated perpendicularly onto the samples mounted on a 6-inch Si wafer. The Ar pressure was kept as $4.8\times10^{-4}$ Torr during milling. The milling rates of LAO layers were calibrated by an in situ secondary ion mass spectroscopy setup, which were 1.7 and 0.8 {\AA}/s for amorphous LAO layer and crystalline LAO layer, respectively.

Figure 1(a) shows a cross-section transmission electron microscopy image of a 20 nm amorphous LAO film grown on an untreated STO substrate at room temperature and 10$^{-6}$ Torr oxygen partial pressure. The LAO layer is seen to uniformly cover the STO substrate. The zoom-in image of an interface region in the inset of Fig. 1(a) demonstrates the amorphous feature of the LAO overlayer, with one or two oriented layers at the interface, which confirms the room-temperature amorphous growth of LAO films on crystalline STO substrates. The first monolayer of the LAO is well oriented, while subsequent layers are increasingly disordered.

The photoluminescence (PL) spectra (excited by a 325 nm laser) of an as-received STO substrate and 20 nm amorphous LAO films deposited on STO substrates at different oxygen partial pressures ranging from 10$^{-1}$ to 10$^{-6}$ Torr are shown in Fig. 1(b). Although the PL intensity of the as-received STO substrate is weak, the characteristic PL peaks of oxygen vacancies at wavelengths ranging from 380 to 420 nm [29,34] in STO can still be seen. The PL intensity of amorphous LAO/STO heterostructures is enhanced by a factor of five to nine relative to the as-received STO substrate, depending on oxygen partial pressure. Moreover, the multiple PL emission peaks are much more pronounced, and the PL intensity increases with decreasing oxygen partial pressure. Considering that 20 nm amorphous LAO films grown on Si substrates present no PL signal [35] and the PL peaks from various defects in LAO bulk crystals appear only at $\sim$600 nm and above [36], we are able to attribute the large enhancement of PL intensity between 350 and 475 nm in amorphous LAO/STO heterostructures to the creation of oxygen vacancies in the STO substrates near their interface during deposition.

\begin{figure*}
\includegraphics[width=6.75 in]{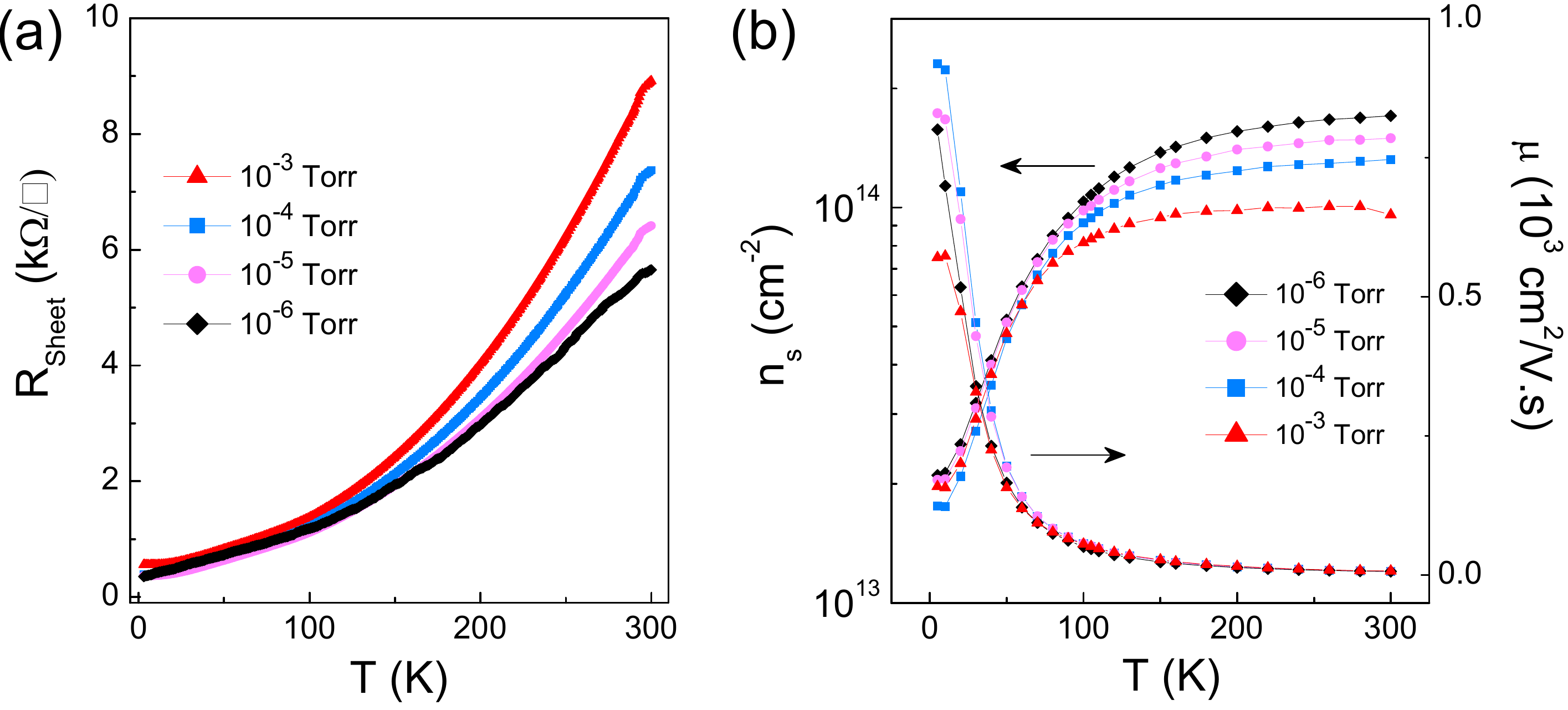}
\caption{\label{fig2} Electrical transport properties of amorphous LAO/STO heterostructures. (a) The temperature dependence of sheet resistance (\emph{R$_{s}$-T}) and (b) sheet carrier density (\emph{n$_{s}$}-T) and the corresponding mobility for 20 nm amorphous LAO/STO heterostructures fabricated at different oxygen pressures from 10$^{-3}$ Torr to 10$^{-6}$ Torr. }
\end{figure*}

The temperature-dependent sheet resistance (\emph{R$_{s}$-T}) of the 20 nm amorphous LAO/STO heterostructures fabricated in different oxygen partial pressures from 10$^{-3}$ to 10$^{-6}$ Torr is shown in Fig. 2(a). As can be seen, the heterostructures exhibit metallic behavior in the whole temperature range. The room temperature sheet resistance increases with oxygen partial pressure. The corresponding carrier density and mobility data are illustrated in Fig. 2(b). The room-temperature carrier density of $\sim$10$^{14}$ cm$^{-2}$ is comparable to that of unannealed crystalline LAO/STO heterostructures [8,19], which were directly cooled down to room temperature in the deposition oxygen pressure after high temperature growth. Moreover, both the temperature-dependent carrier density (\emph{n$_{s}$-T}) and mobility of such amorphous LAO/STO heterostructures are similar to those of unannealed crystalline LAO/STO heterostructures [5,8,19], including the carrier freeze-out effect below $\sim$100 K.

To examine the conductivity and band gap of amorphous LAO, we deposited 150 nm amorphous LAO films on large band gap substrates MgO and Al$_{2}$O$_{3}$. By electrical and ultraviolet-visible spectroscopy measurements, it was found that amorphous LAO is highly insulating with a band gap greater than 5 eV [35], similar to crystalline bulk LAO. Furthermore, any possible built-in potential in amorphous LAO-STO heterostructures should be negligible and only confined to the first or two quasi-crystalline layers of LAO at the interface. The PL spectra in Fig. 1(b) indicate the presence of oxygen vacancies in STO substrates. We are therefore led to conclude that the conductivity emerging at the interface between amorphous LAO films and STO substrates originates largely from oxygen vacancies created in STO near the interface during film deposition.

The \emph{R$_{s}$-T} curve of a 20 nm amorphous LAO/STO heterostructure [35] fabricated at 10$^{-2}$ Torr behaves differently from other heterostructures fabricated at lower oxygen pressure. Indeed, there is a sheet resistance minimum at $\sim$18 K, which was also observed in crystalline LAO/STO heterostructures [8]. Moreover, the sheet resistance of an amorphous LAO/STO heterostructure can be tuned by a back gate voltage. The large tunability in the sheet resistance (60\% variation between $\pm$60 V) and in the MR [35] is comparable to the electric field effect in a crystalline heterostructure [17].

\begin{figure}
\includegraphics[width=3.375 in]{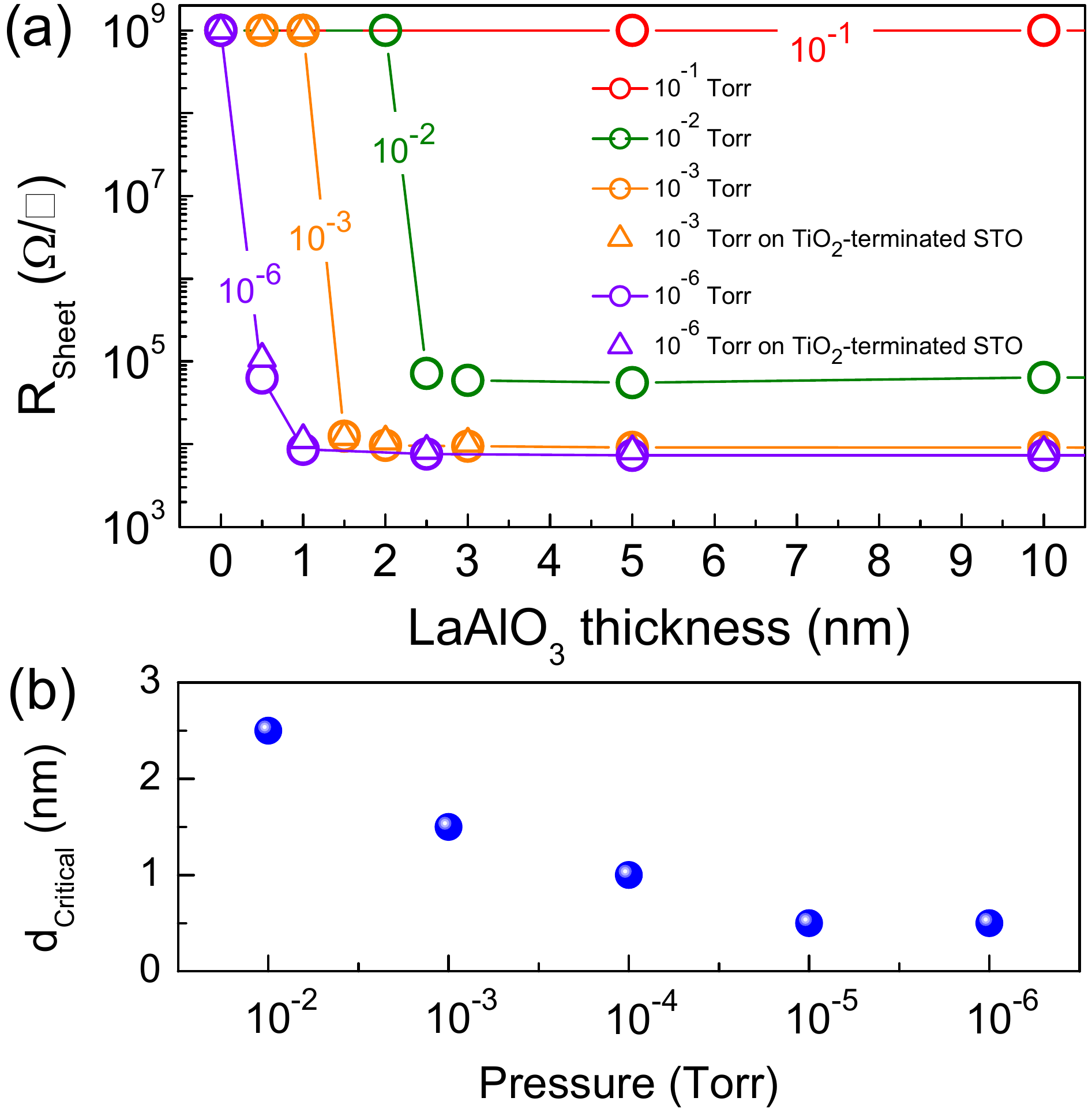}
\caption{\label{fig3} Critical thickness for appearance of conductivity in amorphous LAO/STO heterostructures. (a) Thickness dependence of room-temperature sheet resistance of amorphous LAO/STO heterostructures prepared at different oxygen pressures and on different STO substrates. Triangle symbols represent TiO$_{2}$-terminated STO substrates, while circles represent untreated STO substrates. (b) Critical thickness as a function of deposition oxygen pressure.}
\end{figure}

We systematically examined the LAO layer thickness dependence of sheet resistance for amorphous LAO/STO heterostructures fabricated in different oxygen partial pressures ranging from 10$^{-1}$ to 10$^{-6}$ Torr. For samples deposited at 10$^{-1}$ Torr, no measurable conductivity was ever detected up to a 100 nm LAO layer thickness. As shown in Fig. 3(a), for samples prepared at 10$^{-2}$ Torr and lower pressure, a sharp drop by more than four orders of magnitude in sheet resistance occurs at a certain LAO layer thickness, which strongly depends on oxygen pressure [Fig. 3(b)]. A similar sharp transition in resistance as a function of overlayer thickness was observed by Chen \emph{et al}. [31] and Lee \emph{et al}. [32]. However, the critical thickness for different pressures in our case is different from those reported by Chen \emph{et al}. This is in contrast to the oxygen-annealed crystalline LAO/STO case, where the critical thickness of 4 unit cells (uc) [7] is robust over a large oxygen pressure range from 10$^{-2}$ to 10$^{-5}$ Torr. We also found that the critical thickness in the amorphous case depends on some other factors such as laser energy and substrate-target distance. For example, as we increased the substrate-target distance by a factor of two and lowered the laser fluence from 1.3 to 0.7 J/cm$^{2}$, the critical thickness at 10$^{-3}$ Torr changed from 1.5 nm to 6 nm. There is no pronounced difference in the sheet resistance when the amorphous heterostructures are fabricated on TiO$_{2}$-terminated rather than on randomly-terminated STO.	

\begin{figure*}
\includegraphics[width=6.75 in]{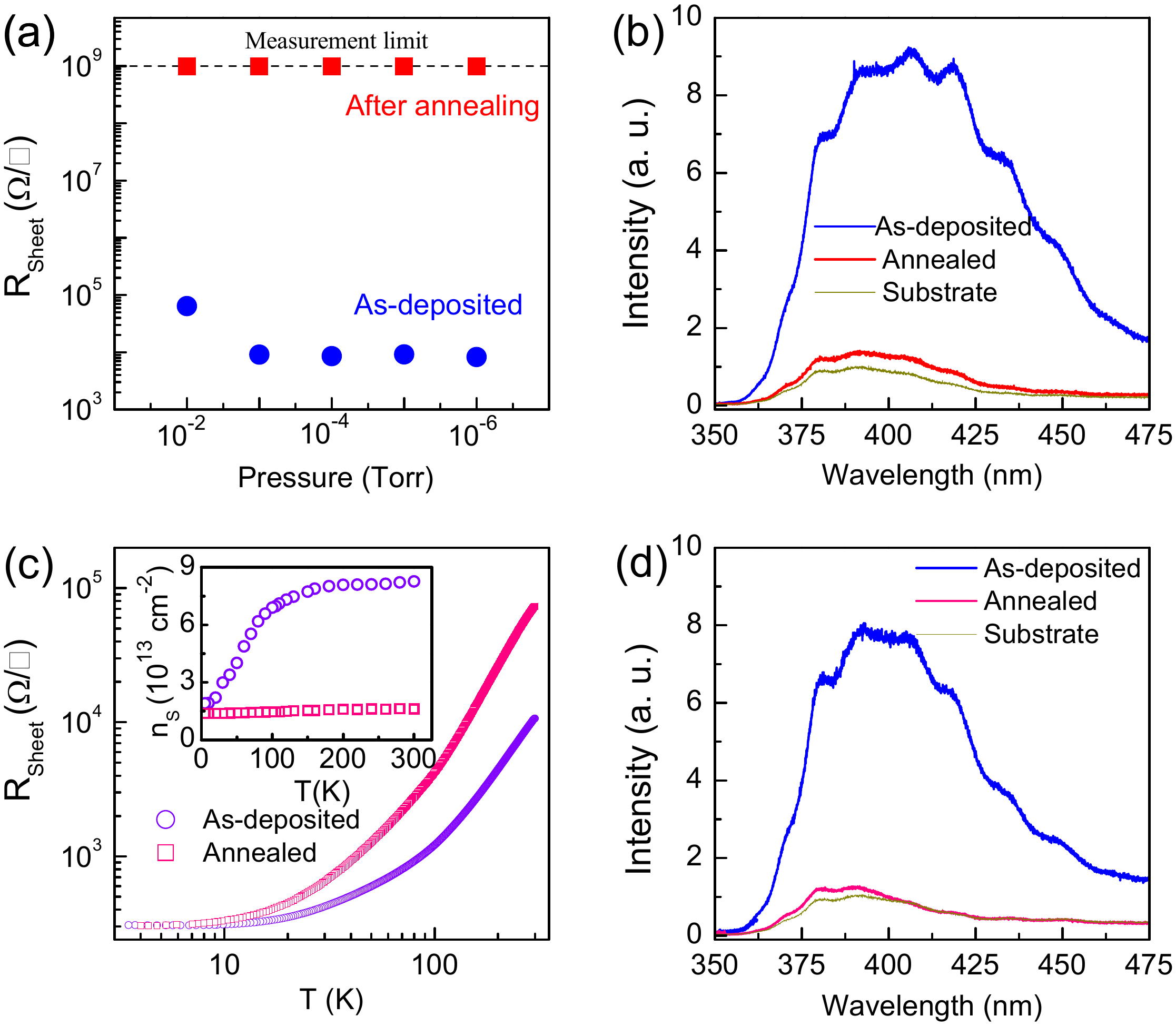}
\caption{\label{fig4} Oxygen-annealing effect. (a) Room-temperature sheet resistance of 20 nm amorphous LAO/STO heterostructures prepared at different oxygen pressures before and after oxygen-annealing in 1 bar of oxygen gas flow at 600 $^\circ$C for 1 h. (b) PL intensity of the 20 nm amorphous LAO/STO heterostructures fabricated at 10$^{-6}$ Torr before and after oxygen-annealing. (c) \emph{R$_{s}$-T}, (inset) \emph{n$_{s}$-T} and (d) PL spectra of a 10 unit cell (uc) crystalline LAO/STO heterostructure prepared at 10$^{-3}$ Torr and 750 $^\circ$C before and after oxygen-annealing in 1 bar of oxygen gas flow at 600 $^\circ$C for 1 h. }
\end{figure*}

Furthermore, the conductivity of all the amorphous LAO/STO heterostructures vanishes after a 1 h post-anneal at 600 $^\circ$C in flowing oxygen (1 bar), as can be seen in Fig. 4(a). At the same time, the PL intensity of all oxygen-annealed amorphous LAO/STO heterostructures decreases significantly and approaches the intensity of the as-received substrate [Fig. 4(b)]. This confirms that oxygen vacancies in STO create the conductivity. To compare the amorphous and crystalline LAO/STO heterostructures, 10 uc crystalline LAO films were grown on TiO$_{2}$-terminated STO substrates at 750 $^\circ$C and 10$^{-3}$ Torr, and then \emph{ex situ} annealed in 1 bar of oxygen gas flow at 600 $^\circ$C for 1 h. They remain conductive, although there is a decrease in carrier concentration and the room-temperature sheet resistance increases by a factor of seven [Fig. 4(c)]; for example, an unannealed crystalline sample has a room temperature carrier density of $8.26\times10^{13}$ cm$^{-2}$, which decreases to $1.62\times10^{13}$ cm$^{-2}$ after post-annealing. Moreover, the \emph{n$_{s}$-T} of the unannealed crystalline LAO/STO sample shows carrier freeze-out below about 100 K, with \emph{n$_{s}$} dropping to $1.90\times10^{13}$ cm$^{-2}$ at 5 K. In contrast, the carrier density of the post-annealed crystalline sample exhibits little temperature dependence, changing from $1.62\times10^{13}$ cm$^{-2}$ at 300 K to $1.38\times10^{13}$ cm$^{-2}$ at 5 K. Such post-annealing experiments are reproducible [35]. The carrier freeze-out effect in unannealed crystalline LAO/STO samples, which also exists in oxygen-deficient STO films [29], is characterized by an activation energy $\epsilon$ of 4.2 meV (fitted from
$n_{s}\propto e^{(-\epsilon/k_B T)}$). In contrast, the activation energy of carriers in oxygen-annealed crystalline LAO/STO samples is much smaller, 0.5 meV. As shown in Fig. 4(d), the PL intensity of the unannealed crystalline sample is greatly enhanced compared to that of its TiO$_{2}$-terminated STO substrate, which reveals the creation of a substantial amount of oxygen vacancies here too during deposition. After post-annealing, the PL intensity falls back to the substrate level, similar to the effect of post-annealing on the PL signal of amorphous samples.

We conclude at this point that oxygen vacancies contribute significantly to the conductivity in both amorphous and unannealed crystalline LAO/STO heterostructures. Specifically, for amorphous LAO/STO samples, the existence of oxygen vacancies in STO substrates is the principal origin of the interface conductivity. For unannealed crystalline LAO/STO samples, oxygen vacancies are only partially responsible for a part of the interface conductivity, which can be eliminated by oxygen-annealing.

\begin{figure}
\includegraphics[width=3.375 in]{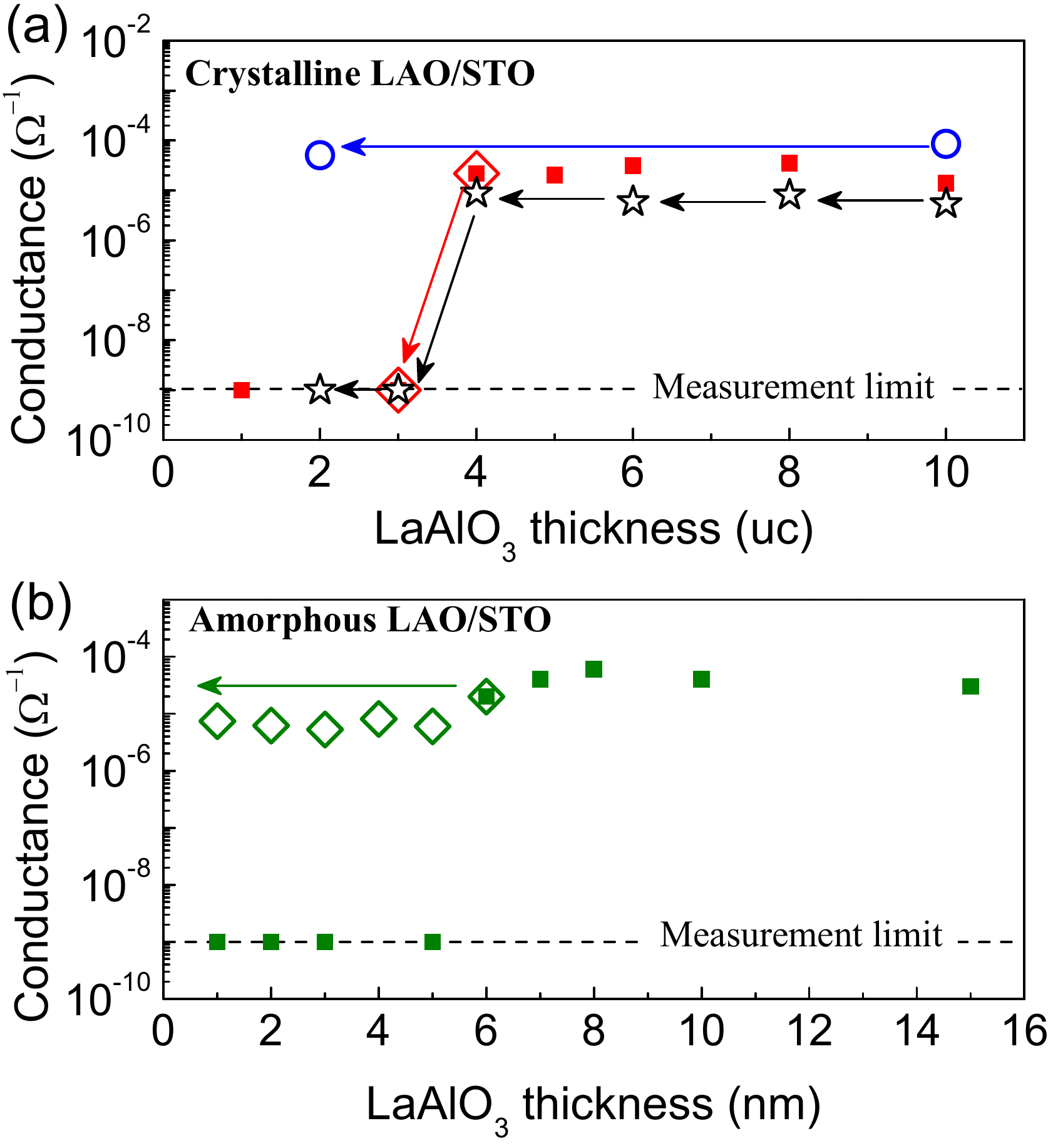}
\caption{\label{fig5} Ar-milling effect. (a) Thickness dependence (red solid squares) of room-temperature conductance of oxygen-annealed crystalline LAO/STO heterostructures fabricated at 10$^{-3}$ Torr and 750 $^\circ$C, showing a critical thickness of 4 uc. The red hollow diamonds denote that the conductivity of the 4 uc sample disappears after the removal of the top 1 uc LAO by Ar-milling. Moreover, the blue hollow circles represent the conductance of an unannealed 10 uc crystalline LAO/STO heterostructure and after the removal of the top 8 uc LAO by Ar-milling. The black hollow stars represent the conductance of another oxygen-annealed 10 uc crystalline LAO/STO sample after step-by-step Ar milling. (b) Thickness dependence (green solid squares) of room-temperature conductance of amorphous LAO/STO heterostructures fabricated at 10$^{-3}$ Torr, showing a critical thickness of 6 nm. The green hollow diamonds represent the conductivity of the 6 nm sample that remains after the removal of the top LAO layer 1 nm at a time by Ar-milling. All the arrows represent the Ar-milling process.}
\end{figure}

To further explore the different mechanisms responsible for the interface conductivity in amorphous and oxygen-annealed crystalline LAO/STO heterostructures, we performed Ar-milling experiments for both types, fabricated at 10$^{-3}$ Torr. Figure 5(a) shows a typical thickness dependence of the conductivity (solid squares) for oxygen-annealed crystalline heterostructures with a critical thickness of 4 uc. We then used a 4 uc sample in the Ar-milling experiments. After removing the top unit cell of LAO, the conductivity disappears, as shown by the hollow diamonds in Fig. 5(a). This result agrees with previous reports [16,37]. Moreover, step-by-step Ar milling of an oxygen-annealed 10 uc crystalline LAO/STO sample generates the same critical thickness of 4 uc for maintaining the interface conductivity (hollow stars). On the other hand, Ar-milling the unannealed crystalline heterostructure from 10 uc down to 2 uc produces little change in conductance (open circles), because the conduction is dominated by oxygen vacancies.

To facilitate control of the Ar-milling rate in the amorphous overlayer case, where the milling rate of LAO was more than twice that of the crystalline case, we intentionally arranged the critical thickness of heterostructures to be 6 nm by increasing the substrate-target distance and decreasing the laser fluence. Figure 5(b) illustrates the LAO layer thickness dependence of conductivity (solid squares) for amorphous LAO/STO samples. As the top amorphous LAO layer is removed, one nm at a time, from a 6nm LAO/STO sample, the conductivity of the heterostructures is retained (hollow diamonds). To check the possible effect of the Ar milling on the conductivity of STO single crystals, an insulating 2 nm amorphous LAO/STO sample was used as reference. After the removal of the top 1 nm of amorphous LAO, the heterostructure remains insulating. This proves that no conductivity is created by the Ar-milling process.

The Ar-milling experiment further confirms that the conductivity in amorphous LAO/STO heterostructures originates principally from oxygen vacancies in the STO substrate. However, the appearance of conductivity in oxygen-annealed crystalline LAO/STO samples is reversible across the critical thickness of 4 uc. Hence this must be closely associated with the interface electronic reconstruction due to the potential build-up in the crystalline LAO overlayer [2-4].This is consistent with the electronic reconstruction at the interface and also the built-in electric field in the polar LAO layer observed by Singh-Bhalla \emph{et al}. [38] and Huang \emph{et al.} [39], where Singh-Bhalla \emph{et al}. performed tunneling experiments and Huang \emph{et al.} conducted band alignment mapping across the crystalline interface by cross-section scanning tunneling microscopy.

\begin{figure*}
\includegraphics[width=6.75 in]{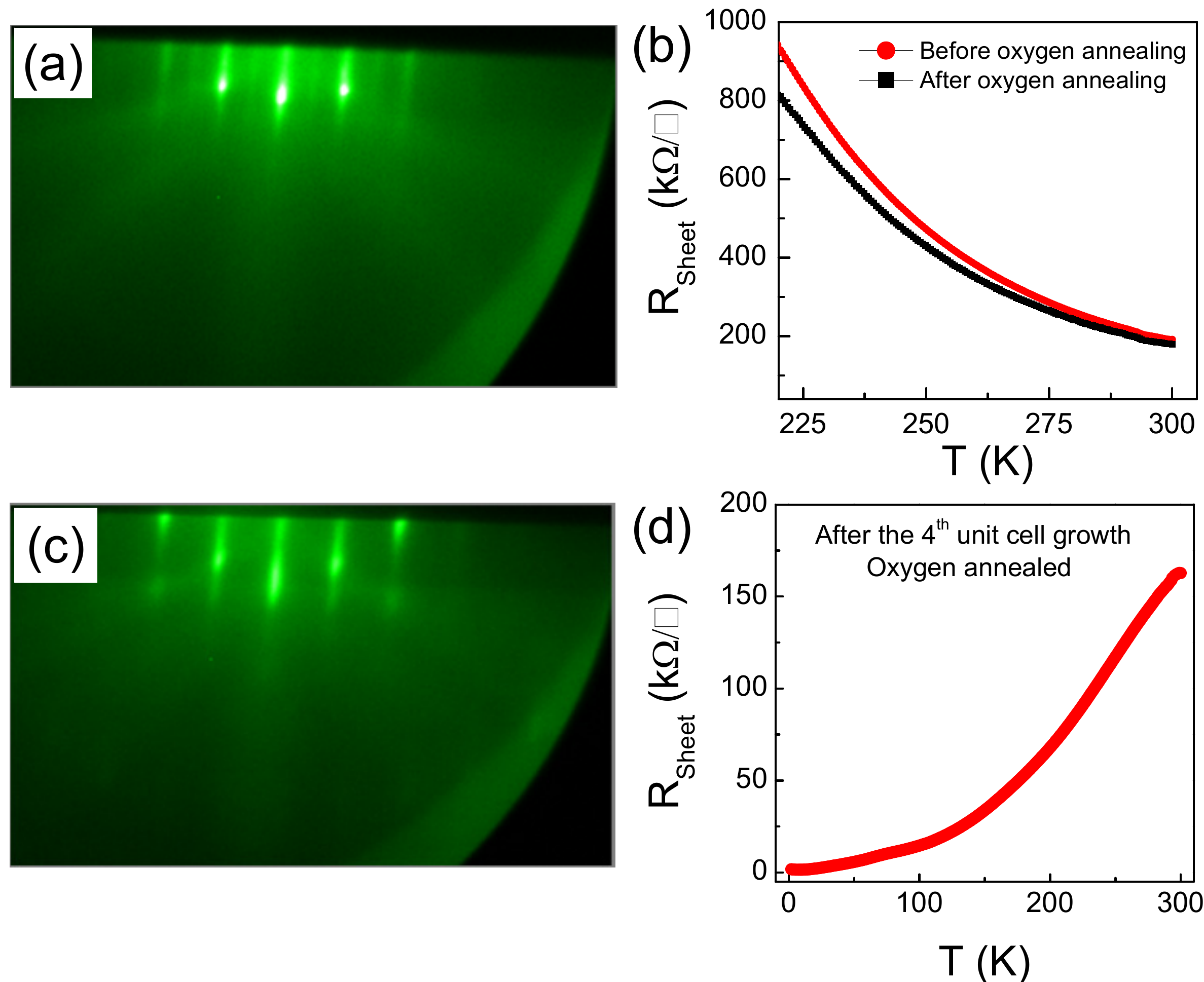}
\caption{\label{fig6} Re-growth experiment. (a) Reflection high energy electron diffraction (RHEED) pattern after depositing a new LAO layer (estimated to be 2 uc) on a crystalline LAO/STO heterostructure with the LAO layer etched from 4 uc to 3 uc. (b) \emph{R$_{s}$-T} curves of the re-grown sample stated in (a) before and after oxygen annealing. (c) RHEED pattern after depositing one uc LAO on an as-grown 3 uc crystalline LAO/STO heterostructure. (d) \emph{R$_{s}$-T} curve of the re-grown sample described in (c) after oxygen annealing.}
\end{figure*}

To explore whether the conductivity of a milled crystalline LAO/STO sample can be restored by depositing another LAO layer, we re-grew LAO on a crystalline LAO/STO heterostructure after the LAO layer was milled from 4 to 3 uc. After the Ar milling ($n\times2$) surface reconstruction was observed in RHEED patterns both before and after deposition [Fig. 6(a)]. No periodic RHEED oscillation was seen [35] during the deposition, which was likely caused by the surface reconstruction of LAO after Ar milling. The thickness of the newly deposited LAO layer was estimated to be 2 uc. Although the re-grown sample shows a measurable room temperature sheet resistance of $\sim$190 k$\Omega$/$\square$, its \emph{R$_{s}$-T} curve exhibits a semiconducting behavior as seen in Fig. 6(b). Oxygen annealing in 600 $^\circ$C and 1 bar of oxygen flow for 1 h only results in a slight change in sheet resistance, which proves that the partially restored conductance is not from oxygen vacancies. Instead, the partially restored conductance suggests that the polarization catastrophe mechanism still works for the re-grown sample but is limited by the poor crystallinity of the re-grown LAO layer.

For comparison, another re-growth experiment was performed. We first fabricated a 3 uc crystalline LAO/STO heterostructure, which was \emph{ex situ} measured to be highly insulating. Then one more unit cell of LAO was deposited on such a heterostructure. No surface reconstruction was seen over the entire deposition process [Fig. 6(c)]. During the re-growth, periodic RHEED intensity oscillation was obtained [35]. The re-grown sample was subsequently oxygen-annealed and the \emph{R$_{s}$-T} curve shows a typical metallic behavior [Fig. 6(d)]. Such re-growth experiments demonstrate that the good crystallinity of the LAO layer is crucial for the polarization catastrophe mechanism for the case of crystalline LAO overlayers.

In addition, there are three intriguing features in Fig. 3 that demand further explanation: i) the sharp conductivity transition, ii) the oxygen pressure dependence of the critical thickness, and iii) the saturation of the sheet resistance with the amorphous overlayer thickness. When depositing the overlayer, chemically reactive species such as Al [40] have a strong propensity to attract oxygen ions from the surface of the STO, even at room temperature. The conductivity transition is explained by percolation of the electrons associated with the oxygen vacancies. The wave function of these electrons will be Bohr-like orbitals with radii of a few nm [35], resulting from the large dielectric constant of STO ($\varepsilon$$_{r}$ = 300 at 300 K) and the large effective mass ($\sim$5m$_{e}$) [41]; the percolation carrier density is $\sim$10$^{13}$ cm$^{-2}$ in one monolayer [35]. The oxygen-depletion process from the STO surface will depend on how much reactive oxygen is available in the ambient atmosphere during deposition, thereby explaining why the critical thickness decreases at lower oxygen pressures. At the carrier densities required for percolation, the vacancy concentration at the STO surface is of the order of a few percent, a value that is already high and the further formation of vacancies will be inhibited [42]. This explains the saturation of the sheet resistance.

In conclusion, despite there being a critical overlayer thickness of LAO for appearance of conductivity at the LAO/STO interface for both crystalline and amorphous forms of LAO, the explanation in the two cases is different. Unlike the 4 uc critical thickness for the oxygen-annealed crystalline heterostructures, there is no universal critical thickness when the LAO is amorphous. The critical thickness then depends sensitively on deposition conditions, and oxygen vacancies in the STO substrate account for the interface conductivity. Oxygen vacancies also contribute substantially to the conductivity of crystalline LAO/STO heterostructures which have not been annealed in oxygen post deposition. The reversible thickness dependence of conductivity across the critical thickness of 4 uc in oxygen-annealed crystalline heterostructures indicates that the interface electronic reconstruction due to the potential build-up in LAO overlayers is ultimately responsible for the conductivity in that case. Moreover, our experiments demonstrate that the crystallinity of the LAO layer is crucial for the polarization catastrophe. Moreover, reproducing similar experimental procedures as reported here would be crucial to reveal the origin and mechanism of the recently reported anisotropic 2DEG at the LAO/STO (110) interface [43] and conductivity at the LAO/STO (111) [44] and other STO-based oxide interfaces.

\begin{acknowledgments}
We would like to acknowledge W. E. Pickett for discussions. We thank the National Research Foundation (NRF) Singapore under the Competitive Research Program (CRP) “Tailoring Oxide Electronics by Atomic Control” (Grant No. NRF2008NRF-CRP002-024), the National University of Singapore (NUS) for a cross-faculty grant, and FRC (ARF Grant No.R-144-000-278-112) for financial support.
\end{acknowledgments}


\end{document}